\documentstyle[epsfig,twocolumn,prl,aps]{revtex}
\include{def}
\def\Journal#1#2#3#4{{#1} {#2} (#4) #3}

\def\NIMR{NIM A}
\def\NPA{Nucl. Phys. A}

\def\PLB{Phys. Lett.  B}
\def\PRL{Phys. Rev. Lett.}
\def\PRC{Phys. Rev. C}
\def\PRD{Phys. Rev. D}

\def\ZPA{Z. Phys. A}

\def\be{\begin{equation}}
\def\ee{\end{equation}}
\newcommand{\gsi}        {$\rm^{5}$}
\newcommand{\heidelberg} {$\rm^{7}$}
\newcommand{\clt}        {$\rm^{3}$}
\newcommand{\bucarest}   {$\rm^{1}$}
\newcommand{\zagreb}     {$\rm^{12}$}
\newcommand{\itep}       {$\rm^{8}$}
\newcommand{\budapest}   {$\rm^{2}$}
\newcommand{\korea}      {$\rm^{10}$}
\newcommand{\msu}      {$\rm^{4}$}
\newcommand{\warsaw}     {$\rm^{13}$}
\newcommand{\dresde}     {$\rm^{6}$}
\newcommand{\kur}        {$\rm^{9}$}
\newcommand{\ires}       {$\rm^{11}$}

\begin{document}
\draft

\title{Azimuthal dependence of collective expansion for
symmetric heavy ion collisions}

\author{
G.\,Stoicea\bucarest, M. Petrovici\bucarest$^{\rm ,}$\gsi,
A. Andronic\bucarest$^{\rm ,}$\gsi, N.~Herrmann\heidelberg,
J.P.\,Alard\clt,
Z.\,Basrak\zagreb,
V.\,Barret\clt,
N.\,Bastid\clt,
R.\,\v{C}aplar\zagreb,
P.~Crochet\clt,
P.\,Dupieux\clt,
M.\,D\v{z}elalija\zagreb,
Z.\,Fodor\budapest,
%
%
O.~Hartmann\gsi,
K.D.~Hildenbrand\gsi,
B.\,Hong\korea,
J.\,Kecskemeti\budapest,
Y.J.\,Kim\korea,
M.\,Kirejczyk\warsaw,
M.\,Korolija\zagreb,
R.\,Kotte\dresde,
T.\,Kress\gsi,
A.\,Lebedev\itep,
Y.~Leifels\gsi,
X.\,Lopez\clt,
M.\,Merschmeier\heidelberg,
W.\,Neubert\dresde,
D.\,Pelte\heidelberg,
F.\,Rami\ires,
W.~Reisdorf\gsi,
D.~Sch\"ull\gsi,
Z.\,Seres\budapest,
B.\,Sikora\warsaw,
K.S.\,Sim\korea,
V.\,Simion\bucarest,
K.\,Siwek-Wilczy\'nska\warsaw,
V.\,Smolyankin\itep,
M.\,Stockmeier\heidelberg,
K.~Wi\'{s}niewski\gsi,
D.\,Wohlfarth\dresde,
I.\,Yushmanov\kur,
A.\,Zhilin\itep\\
(FOPI Collaboration)\\
and\\
P.\,Danielewicz\msu$^{\rm ,}$\gsi
}

\address{
\bucarest~National Institute for Nuclear Physics and Engineering, Bucharest, Romania\\
\budapest~Central Research Institute for Physics, Budapest, Hungary\\
\clt~Laboratoire de Physique Corpusculaire, IN2P3/CNRS,
and Universit\'{e} Blaise Pascal, Clermont-Ferrand, France\\
\msu~Michigan State University, East Lansing, USA\\
\gsi~Gesellschaft f\"ur Schwerionenforschung, Darmstadt, Germany\\
\dresde~Forschungszentrum Rossendorf, Dresden, Germany\\
\heidelberg~Physikalisches Institut der Universit\"at Heidelberg, Heidelberg, Germany\\
\itep~Institute for Theoretical and Experimental Physics, Moscow, Russia\\
\kur~Kurchatov Institute, Moscow, Russia \\
\korea~Korea University, Seoul, South Korea\\
\ires~Institut de Recherches Subatomiques, IN2P3-CNRS, Universit\'e
Louis Pasteur, Strasbourg, France \\
\zagreb~Rudjer Boskovic Institute, Zagreb, Croatia\\
\warsaw~Institute of Experimental Physics, Warsaw University, Poland\\
}

\maketitle

\vspace{-0.8cm}

\begin{abstract}
Detailed studies of the azimuthal dependence of the mean
fragment and
flow energies in the
Au+Au and Xe+CsI systems are reported as  a function of incident
energy and centrality.
Comparisons between data and model calculations show
that the flow energy values along different azimuthal directions could be
viewed as snapshots of the fireball expansion with different exposure
times.
For the same number of
participating
nucleons more transversally elongated participant shapes from the heavier
system produce less collective transverse energy. Good
agreement with BUU calculations is obtained for a soft nuclear
equation of state.
\end{abstract}


\vspace{-0.5cm}

 One of the main motivations to study heavy ion collisions at high energy is
to obtain information on the equation of state (EoS) for nuclear matter
under conditions of pressure and temperature different from those in
normal nuclei. The search for hot and dense nuclear matter created in
such collisions is confronted with dynamical consequences of the high
incident energy necessary to reach such conditions and
with the difficulty to reach a thermal equilibrium in finite systems.
Dynamical aspects refer not only to the initial phase of the collision,
but also to the evolution stage of the formed fireball. Thus, detailed
experimental information on the expansion dynamics is required. The
simplest situation corresponds to central collisions with the advantage
of the azimuthal symmetry and of the lack of spectator matter. Predicted
in early seventies \cite{1,2}, the collective expansion of the hot and
dense fireball produced in central collisions was evidenced
experimentally \cite{3,4,5,6,7,8,9,10,11}. Although central collisions
seem to deliver the cleanest signal on the collective expansion on first
sight, two issues are worth mentioning: i) While the axial symmetry of
the dynamical evolution holds, the spherical symmetry has to be
inspected. Preequilibrium emission and transparency effects could
influence the spherical symmetry of the expanding system. ii) With regard
to reaching pressure, the nuclear matter, not being confined in
transverse directions, can escape freely in any direction perpendicular
to the collision axis starting from the very first moments of the
collision. For reduced centrality, other complications appear. One has to
deal with rotating expanding objects in the presence of spectator matter.
Nevertheless there are also some advantages in studying less central
collisions: i) rotation and shadowing can be used as internal clocks for
getting deeper information on the expansion dynamics, ii) the centrality
can be used to control the shape and content of the fireball and of the
shadowing matter, iii) for a given centrality the passage time of the
shadowing objects can be controlled varying the incident energy, iv) the
confinement of the spectators becomes more efficient in transverse
directions than in the central collisions. Symmetry considerations imply
two dominant components in the particle transverse emission: azimuthally
symmetric emission and an elliptic modulation of that emission
(squeeze-out) which has been predicted by hydrodynamical calculations
\cite{12} and extensively studied experimentally
\cite{13,14,15,16,17,18,19,20,21,22,23,24}. The squeeze-out has been
studied, in particular, as a function of centrality, type of emitted
particle, transverse momentum and mass of the colliding systems. A
considerable collectivity was identified in the transverse emission
pattern and it was found, from fitting the energy spectra with a radially
expanding source, that the collectivity itself exhibits an elliptic
modulation \cite{19}. The present paper is devoted to the elliptic
modulation of collectivity as reflected in the dependence of average
fragment energies on the fragment mass \cite{22}.

Notably, without any modulation of collectivity, the
squeeze-out itself could just represent pure geometric
shadowing of particle emission from the participant zone.
Significant modulations, as a
counterpart, should show significant influence by early
pressure, with the compression and excitation energy
getting converted into collective energy at the stage of
high density when the spectators are present.
Comparisons of different centralities and system masses are
needed in order to understand how the geometry of the participant
and spectator zones affects the collective expansion.
Excitation functions for the collectivity anisotropy are
mandatory to understand how the collective expansion builds up at
different densities and excitations.

The present work is based on $^{197}$Au +$^{197}$Au and $^{130}$Xe +
$^{133}$Cs$^{127}$I data from FOPI Phase II experiments at the SIS of
GSI. Detector details can be found elsewhere \cite{25,26,24}. The main
component implemented in the FOPI configuration is a Central Drift
Chamber (CDC) \cite{26}. Two criteria for centrality selection were
combined: the particle multiplicity, with a higher selectivity at large
impact parameters and the ratio of transverse and longitudinal energies
in the c.m.\ system $E_{rat}=\sum_{i}E_{\perp,i}/\sum_{i}E_{\parallel,i}$,
as a better filter for more central collisions. Two windows in the CDC
multiplicity, CM2 and CM3, have been used to select impact parameters in
the range of 6-8 fm and 4-6 fm. Similarly, the ER4 and ER5 windows in
$E_{rat}$ select the ranges 2-4 fm and 0-2 fm, respectively. The impact
parameter estimates and the estimates of the number of participating
nucleons $A_{part}$ are based on the geometrical $"$sharp cut-off$"$
approximation. For reaction plane determination, the transverse-momentum
analysis method has been used \cite{27}.
 Studies of the squeeze-out phenomena revealed from the very beginning
\cite{13} the importance of
performing the azimuthal distribution analysis in a reference frame with the
polar axis along the sidewards flow direction. The present analysis used
a reference frame in which the ellipsoidal pattern of the azimuthal
particle distribution at mid-rapidity maximizes the ratio of
the two transverse
semiaxes \cite{19,22,24}. The previous comprehensive description of
the elliptic modulation of collectivity in the beam range 0.25
- 1.15 A$\cdot$GeV \cite{19} was
based on a parameterization of the deuteron (A=2) tranverse-mass spectra with an
expression
characteristic of radially symmetric shell expansion.
The use of such an expression for studying the modulations
is inherently contradictory and implies a specific model.
We prefer to present experimental
information free from any model.

For the mentioned reasons we concentrate on the mean kinetic
energy in the c.m. system,
$\langle E_{kin}^{cm} \rangle$, and on the flow energy
extracted from its dependence on
the mass of the reaction products (Z=1,2 and 3)
within a polar angular range of $80^{\circ}$$\leq\theta_{cm}\leq$$100^{\circ}$ \cite{22}.
In order to extract
\mbox{$\langle E_{kin}^{cm} \rangle$}, one needs complete energy
spectra.
Due to the peculiar shape of the
shadows of
subdetector borders in the rotated reference frame,
the experimental
spectra have been analyzed  in the azimuthal ranges [0$^\circ$, 90$^\circ$]
and [270$^\circ$, 360$^\circ$].
These two ranges have been overlapped to
decrease statistical errors.
They are plotted as full symbols in Fig. 1 for five bins from $0^\circ$ to
$90^\circ$ and were then reflected (open symbols)
in order to generate the full angular
range of $0^\circ$-$360^\circ$.
\begin{figure}[thb]
\vspace{1.1cm}
\centering{\mbox{\epsfig{file=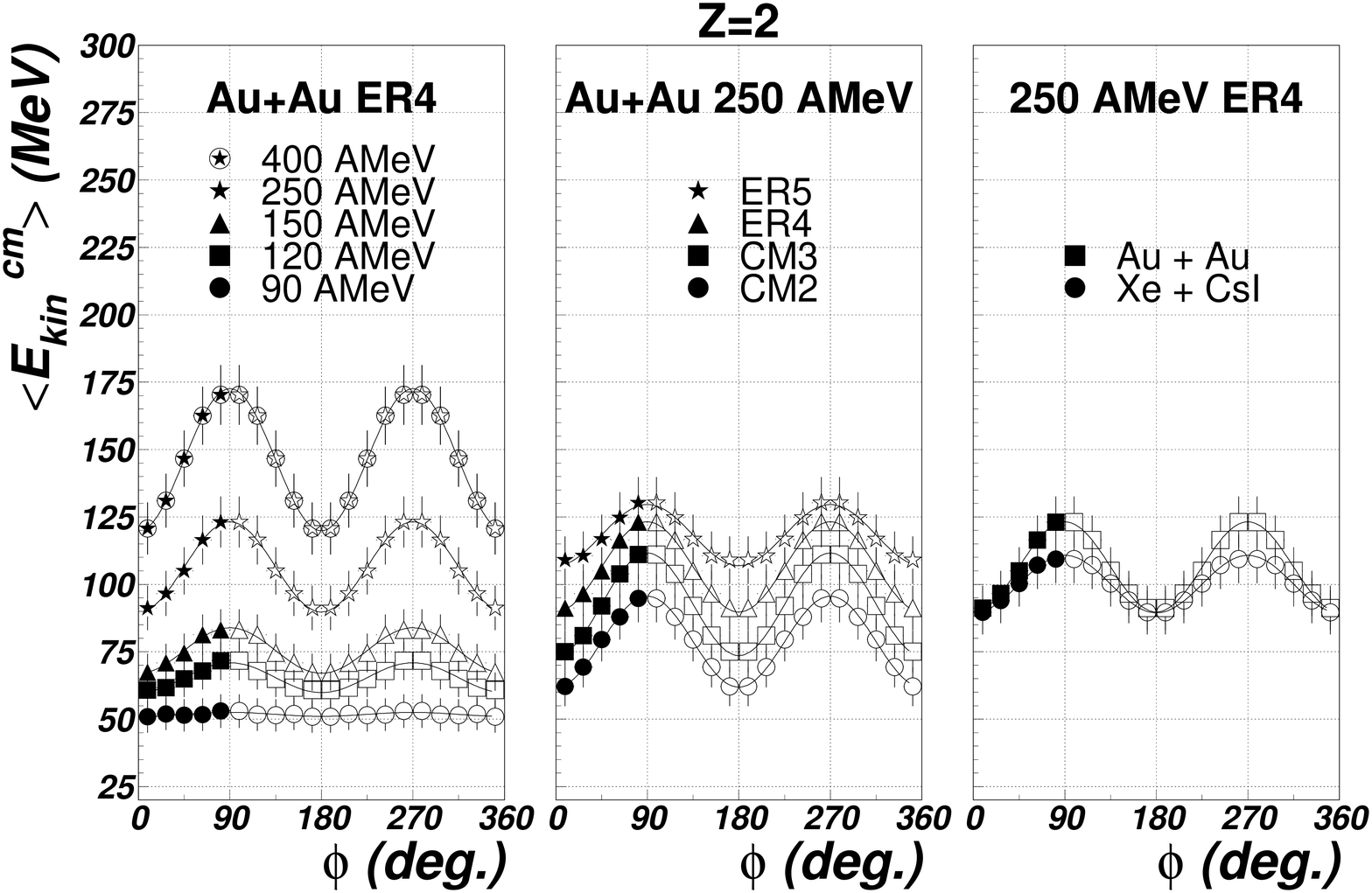, width=.125\textwidth,
bbllx=420pt,bblly=253pt,bburx=700pt,bbury=569pt}}}
\vspace{1.75cm}
\caption{
Mean c.m. energy
$\langle E_{kin}^{cm} \rangle$ of Z=2
products as a function of the azimuthal angle,
shown for:
Au+Au at ER4 centrality and different energies (left);
Au+Au at 250 A$\cdot$MeV and different centralities (middle), and Au+Au
and Xe+CsI at 250
A$\cdot$MeV and ER4 centrality (right). The error bars include systematic effects.}
\label{fig-1}
\end{figure}

\vspace{-.4cm}

 Figure \ref{fig-1} presents, as an example, the azimuthal
dependence of $\langle E_{kin}^{cm} \rangle$
for Z=2 products as a function of the incident energy in
Au + Au at
ER4 centrality, as a function of centrality in Au + Au at 250 A$\cdot$MeV,
and for the two measured systems at 250 A$\cdot$MeV and ER4
centrality.
The presented information is \underline{independent} of the anisotropy
of the yield distribution.
The main contribution to the error bars comes from systematic effects,
the
statistical ones being at the level of symbol sizes. We
had to combine information from two subdetector systems and this was done
by looking at all fragments as a function of their charge. In doing this,
we had to correct the Z=1 and Z=2 energy spectra from the CDC, based on
previous FOPI data \cite{8}, in order to
take into account the fact that the $^3He$ fragments are not well separated
from the tritons in some parts of the momentum space.
The azimuthal asymmetry of $\langle E_{kin}^{cm} \rangle$
increases as a function of incident energy and
mass of the colliding nuclei, reaching a maximum in mid-central
collisions.
The value averaged over the azimuth increases with the beam
energy, centrality and mass of interacting system.
As shown by the solid lines, the data follow very well the behavior
$\langle E_{kin}^{cm} \rangle$ = $E_{kin}^0$ -
$\Delta$$E_{kin}$$\cdot \cos{ 2 \Phi}$. Fits to the dependence
of mean kinetic energies
$\langle E_{kin}^{cm} \rangle$ on fragment mass A
with the nonrelativistic expression

\begin{equation}
\langle E_{kin}^{cm} \rangle \approx\frac{1}{2}
A\cdot m_0<\beta^2_{flow}>+\frac{3}{2}$"$T$"$,
\end{equation}

\noindent
yield the average flow energy per nucleon $E_{coll}$ and the
$"$temperature$"$  $"$T$"$ ($m_0$ is the nucleon rest mass).
Lack of an explicit treatment of the Coulomb
effects leads to a systematically overestimated value
of the real temperature T \cite{8}.

\vspace{-.3cm}
\begin{figure}[thb]
\begin{minipage}[b]{0.47\linewidth}
\centering\epsfig{file=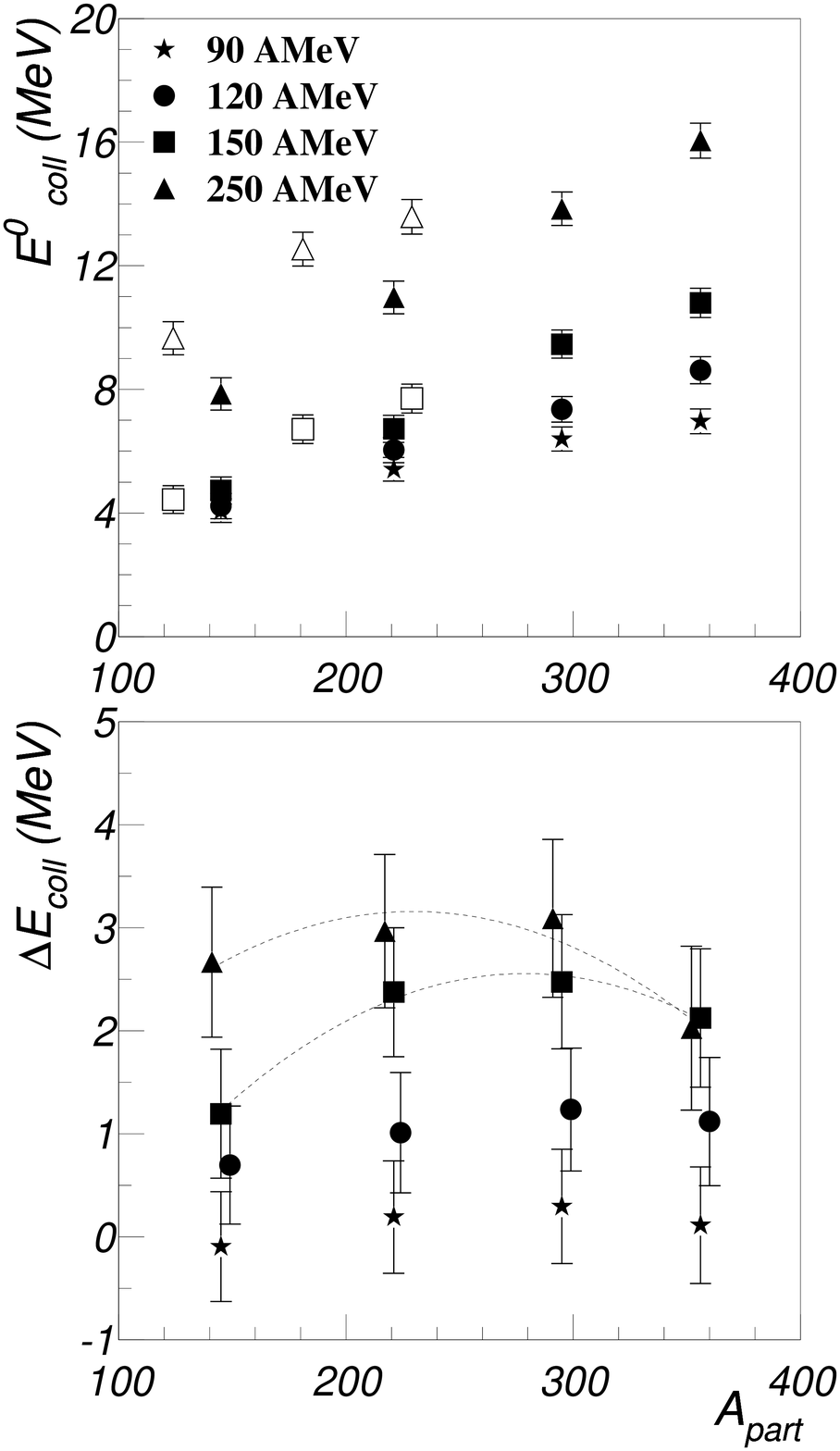, width=\linewidth}
\end{minipage}\hfill
\begin{minipage}[b]{0.47\linewidth}
\centering\epsfig{file=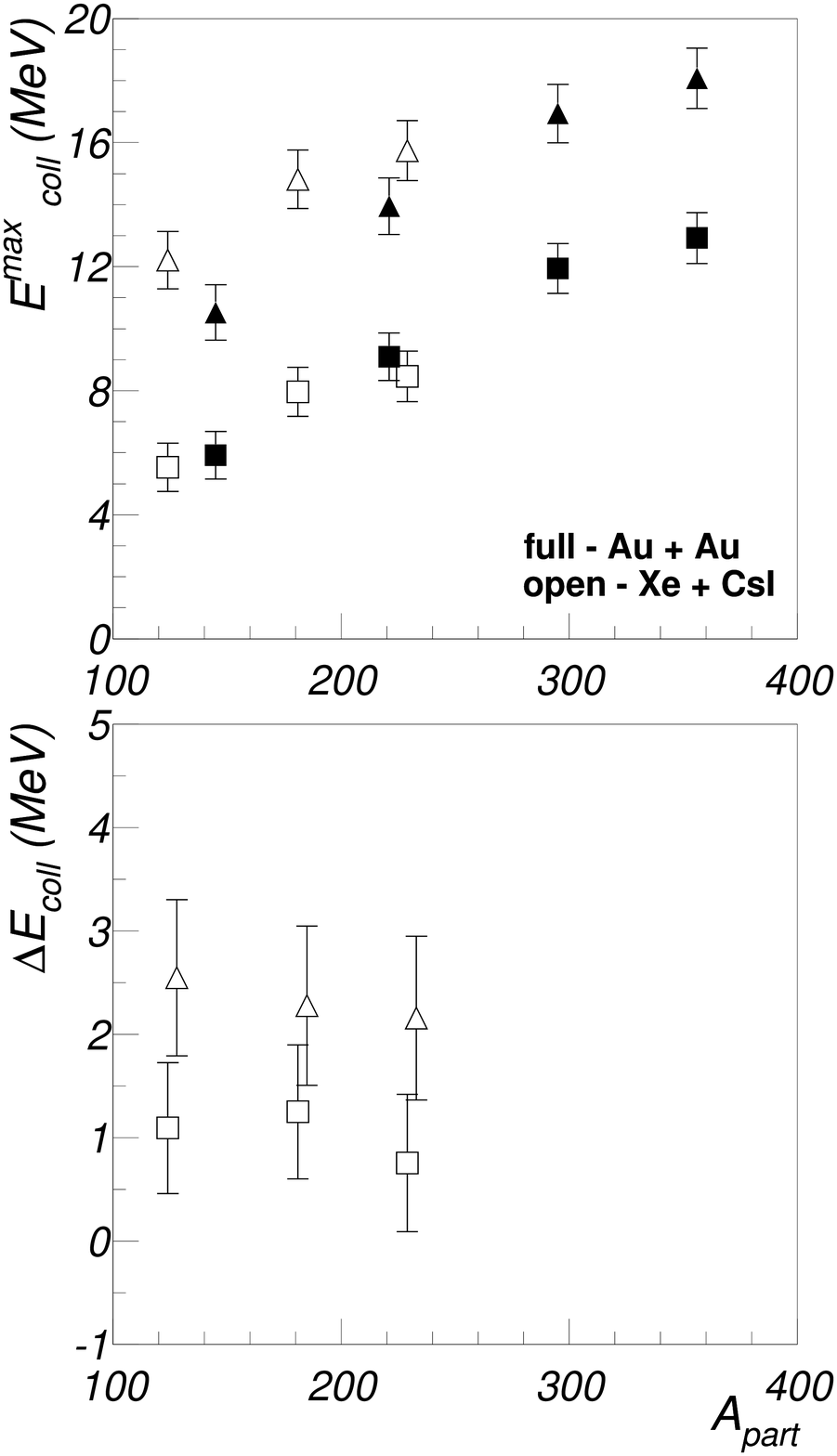, width=\linewidth}
\end{minipage}\hfill
\caption{$E_{coll}^0$ (top-left panel),
$E_{coll}^{max}$ (top-right panel)
and $\Delta$$E_{coll}$
(bottom)
as a function of $A_{part}$, corresponding to the four centralities
mentioned in the text, at different incident energies in
the Au+Au (full symbols) and Xe+CsI (open symbols) systems.
Second-order polynomial fits, represented by dashed lines,
serve to guide the eye.
} \label{fig-2}
\end{figure}

\vspace{-0.3cm}
The flow energy $E_{coll}$=1/2$\cdot$$m_0$$<\beta^2_{flow}>$ and $"$T$"$
can be fit nicely with
$E_{coll}$ = $E_{coll}^0$ -
$\Delta$$E_{coll}$$\cdot \cos{2\Phi}$
and
$"$T$"$ = $"$$T_0$$"$ -
$\Delta$$"$T$"$$\cdot \cos{2\Phi}$,
respectively.
$E_{coll}$ exhibits a strong elliptic anisotropy,
with the largest energy values in the direction
perpendicular to the reaction plane.  Both the average over the azimuth
of the collective energy
$E_{coll}^0$ and the elliptic anisotropy $\Delta$$E_{coll}$
increase continuously with the incident
energy over the studied region.
The temperature parameter $"$$T$$"$ stays roughly
constant as a function of the azimuth at three lowest beam
energies and develops
oscillations $\Delta$$"$T$"$ in the range 1-2.5 MeV, at
250 A$\cdot$MeV.  These could be indicative of the small
variations of temperatures at the different sides
of the participant fireball \cite{7}. Figure
\ref{fig-2} provides global
results:
$E_{coll}^0$, $E_{coll}^{max}$($E_{coll}$ at $90^\circ$) and
$\Delta$$E_{coll}$
as functions of $A_{part}$ for both measured systems.
At 90 A$\cdot$MeV, the in-plane and out-of-plane flow values are
very similar
($\Delta$$E_{coll}$$\sim$0)
 at all centralities in Au+Au,
paralleling the observations for the yields for which a
transition
from in-plane to out-of-plane enhancement was found \cite{24}
as a function of incident energy.
At all energies, $E_{coll}^0$ increases with
the centrality corresponding to an increasing
baryonic number $A_{part}$ of the fireball.
Although the error bars, which include the systematic effects, are
large, the relative errors for the data at different $A_{part}$
are small,
of
the order of 0.08 MeV. At higher energies, a maximum of $\Delta$$E_{coll}$
in mid-central Au+Au collisions becomes visible.
A comparison of the two systems 250 A$\cdot$MeV in the ~$90^\circ$
direction, outside of the spectator shadow,
Fig.\ 2 top-right, shows a difference of the
order of 20\% in $E_{coll}^{max}$ between the two systems, at
the same $A_{part}$.
One should notice that, for the same $A_{part}$,
the fireball produced in the Xe + CsI case is more
spherical than that in the Au + Au case.
This remarkable dependence of the transverse flow energy on the shape of the
expanding nuclear zone is reported here for the first time.
Flatter
fireballs yield on the average less transverse expansion than
more spherical ones.

\vspace{-0.3cm}
\begin{figure}[thb]
\centering\mbox{\epsfig{file=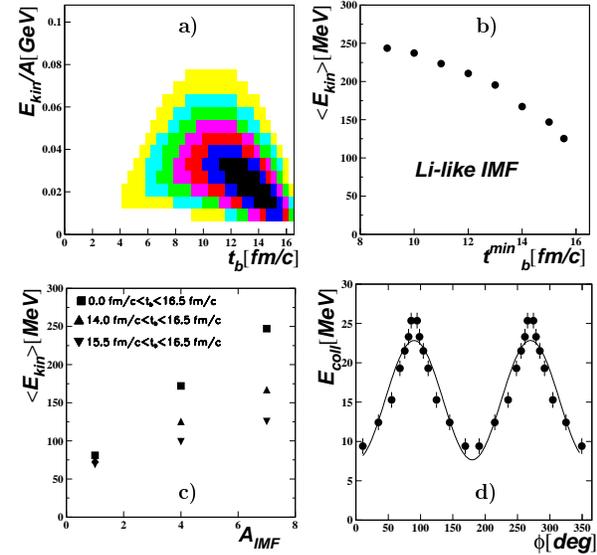,
width=0.47\textwidth}}
\vspace{0.25cm}
\caption{a) Li-like fragment yield distribution in the plane of
$E_{kin}$ vs break-up
time $t_b$ in the hybrid model.  b)
Mean kinetic energy of Li-like
fragments emitted after a break-up time
$t^{min}_b$.
c) Mean kinetic energy of Z = 1, 2 and 3 fragments as a
function of their mass for different ranges of the
break-up times.
d) Collective energy as a function of azimuthal angle in the
hybrid model. }
\label{fig-3}
\end{figure}

\vspace{-0.3cm}
To get an insight on the main mechanism behind the observed experimental
trends, we have calculated the expansion dynamics
within a semi-analytical
hybrid model \cite{7}
for a 200-nucleon fireball, which roughly corresponds to
$A_{part}$
at CM3 centrality in Au + Au collision.  Notably, this model, in which
expansion dynamics is combined with statistical features
of cluster formation at freezeout, predicts a decrease
of the collective
energy and of the temperature as the system deexcitation progresses.
A two dimensional
yield distribution as a function of
$E_{kin}$ and
break-up time $t_b$
for Li fragments in 250 A$\cdot$MeV Au+Au collision is
presented in Fig.\ 3a.  In the directions outside of
the spectator shadow, i.e.\ perpendicular
to the reaction
plane, fragments with kinetic energies corresponding to all
break-up times will be
visible.  The situation changes if the observer views the
reaction from the reaction plane.
Particles emitted directly from the fireball are visible once the spectators
moved
apart from the collision zone by a distance corresponding to
at least half of the
transition time.  Those particles which are emitted earlier will get redirected.
At later
times, the $E_{kin}$ distributions become narrower following the
decompression and this effect gets more
pronounced for heavier fragments.
Figure 3b shows the mean energy of Li-like fragments
as a function of a starting time $t^{min}_b$ for emission, from
integrating over the yield in panel a).
Mean kinetic energies as a function of fragment mass
number A, for different emission
intervals in $t_b$, can be seen in Fig.\ 3c.  If one
calculates the emission in a simple geometric picture, under
the assumption that the centers of original nuclei pass each
other in the middle of the expansion time shown in Fig.\ 3a,
one gets the behavior of flow energies represented in Fig.\ 3d.
Although quite
simplistic, the model nicely reproduces the qualitative trends seen in the
data, showing that the flow energy values could be viewed as
snapshots of the fireball expansion dynamics with different
exposure times for different azimuthal directions.

\vspace{-0.4cm}
\begin{figure}[thb]
\begin{center}
{\epsfig{file=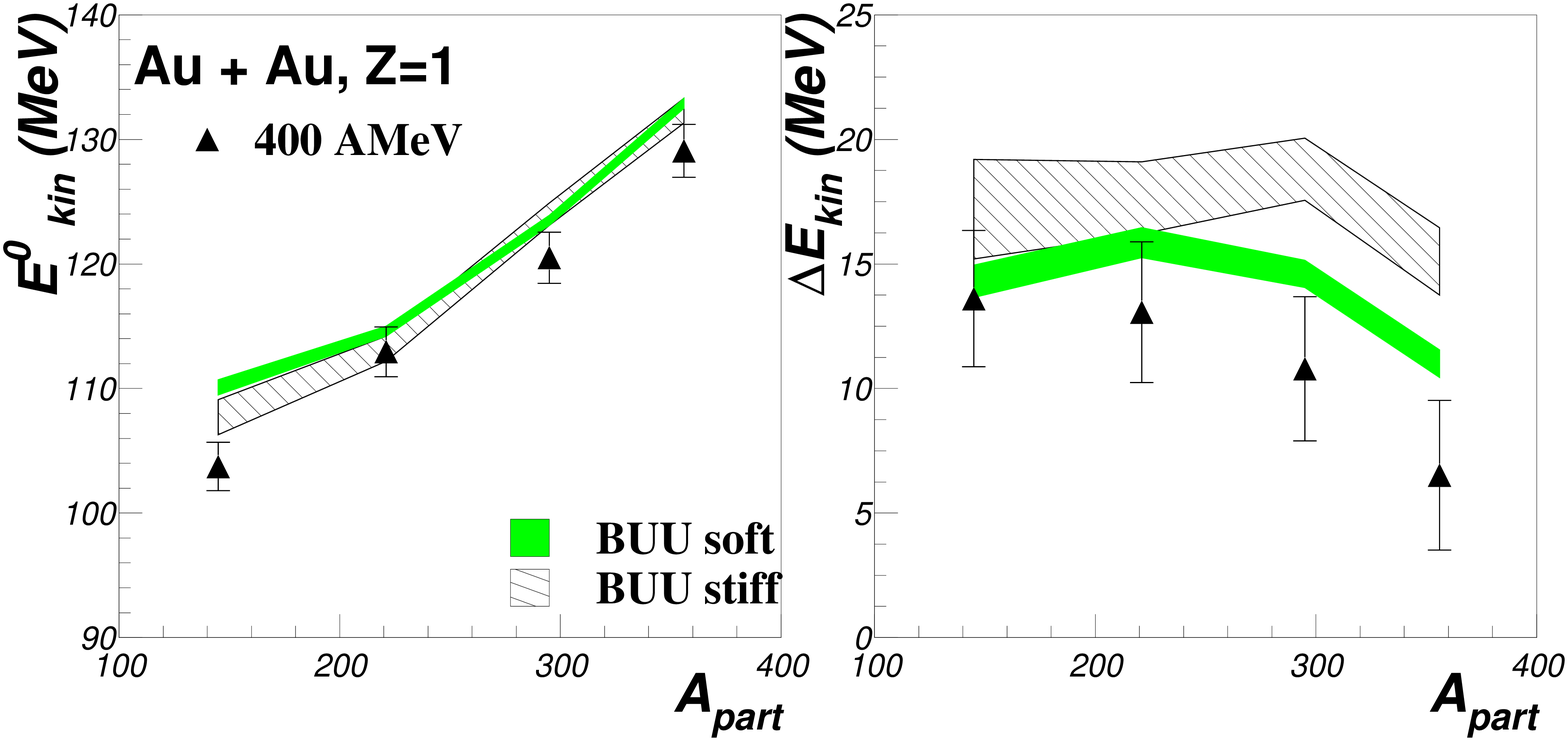, width=.47\textwidth}}
\end{center}
\vspace{-0.2cm}
\caption{
$E_{kin}^{0}$ and $\Delta$$E_{kin}$
as a function of $A_{part}$, for Z=1 (A=1,2,3) fragments,
Au+Au at 400 A$\cdot$MeV. The experimental
results are represented by triangles, while the BUU results are
represented by gray zones for soft EoS and by dashed zones for stiff
EoS, respectively.
}
\label{fig-4}
\end{figure}

\vspace{-0.4cm}
It is obvious that preequibrium processes can be important
and that dynamic long-term anisotropies can be produced in the
collective expansion with either effect being absent in the simple
scenario above.  Thus, a comparison with the {\em ab initio}
microscopic transport model becomes important.
Correspondingly, for the 400
A$\cdot$MeV
Au+Au collisions, we present in Fig.\ 4 the results of the
Boltzmann-Uehling-Uhlenbeck (BUU) transport code \cite{28}
using momentum dependent mean fields ($m^*$/m=0.79), in-medium
elastic cross sections
($\sigma = \sigma_0 \,
\mbox{tanh}(\sigma^{free}/\sigma_0)$ with $\sigma_0 =
\rho^{-2/3}$)
and soft (K=210 MeV, gray zone) or
stiff (K=380 MeV, dashed zone) EoS. The light fragments (up to A=3) are
produced
in a few-nucleon processes inverse to composite break-up.
The measured relative yields
are
nicely reproduced, especially at higher incident energies \cite{8}.
The
calculated yields have been smeared according to the measured
reaction-plane dispersion.
For
$\langle E_{kin} \rangle$, little sensitivity to EoS is found,
with either EoS parameterization yielding a quite good agreement
with the data.
As far as the
values of $\Delta E_{kin}$ are concerned, the calculations
with the soft EoS reproduce the overall trend of the experiment
while the calculations with the stiff EoS overestimate
significantly especially at higher centralities (lower impact parameter)
the measured values.

 In summary, we presented results on the azimuthal dependence
of mean fragment and flow energies in two symmetric
systems, for different centralities and incident energies.
Corroborated by
model estimates, indications emerged that different
regions of the azimuth capture different periods of the central
fireball expansion.  In comparing results from two symmetric
systems, it was possible to evidence that a more spherical
fireball produces more transverse flow at a given participant
number than a deformed fireball.
Comparisons with transport
code predictions demonstrated that the soft EoS, in combination
with a momentum-dependent mean fields and with in-medium
cross sections,
gives a good agreement with the
experiment.

This work was partly supported by the German BMBF under contract
RUM-99/010 and 06HD953, by the NSF under Grant PHY-0245009 and by KRF
under Grant 2002-015-CS0009.

\end{document}